# Correlation versus randomization of jerky flow in an AlMgScZr alloy using acoustic emission

T.A. Lebedkina,[1,2] D.A. Zhemchuzhnikova,[1] M.A. Lebyodkin[1*]

[1] *Laboratoire d'Etude des Microstructures et de Mécanique des Matériaux (LEM3), Université de Lorraine, CNRS, Arts & Métiers ParisTech, 7 rue Félix Savart, Metz, France*

[2] *Togliatti State University, Belorusskaya St. 14, Tolyatti 445020, Russia*

**Abstract**
Jerky flow in solids results from collective dynamics of dislocations which gives rise to serrated deformation curves and a complex evolution of the strain heterogeneity. A rich example of this phenomenon is the Portevin–Le Chatelier effect in alloys. The corresponding spatiotemporal patterns showed some universal features which provided a basis for a well-known phenomenological classification. Recent studies revealed peculiar features in both the stress serration sequences and the kinematics of deformation bands in Al-based alloys containing fine microstructure elements, such as nanosize precipitates and/or submicron grains. In the present work, jerky flow of an AlMgScZr alloy is studied using statistical analysis of stress serrations and the accompanying acoustic emission. As in the case of coarse-grained binary AlMg alloys, the amplitude distributions of acoustic events obey a power-law scaling which is usually considered as evidence of avalanche-like dynamics. However, the scaling exponents display specific dependences on the strain and strain-rate for the investigated materials. The observed effects bear evidence to a competition between the phenomena of synchronization and randomization of dislocation avalanches, which may shed light on the mechanisms leading to a high variety of jerky flow patterns observed in applied alloys.

* Corresponding author. *E-mail* address: mikhail.lebedkin@univ-lorraine.fr

## 1. Introduction

The interest in jerky flow in dilute alloys, known as the Portevin–Le Chatelier (PLC) effect, has not diminished since its discovery [1]. One reason for such sustaining interest is that the PLC effect is undesirable in practice because of its detrimental effect on the material formability. On the other hand, it is a striking example of self-organization of dislocations. The complex collective nature of plastic deformation processes is now generally recognized for both crystalline [2-6] and amorphous [7] solids. However, whereas jerkiness manifests itself directly on the deformation curves of micropillars [6], it is usually smoothened out in the case of bulk samples, except for several mechanisms leading to macroscopic plastic instabilities (see, e.g., the reviews in Refs. [4,5]). Among these mechanisms, the PLC effect displays the richest behavior unifying many kinds of collective dynamics observed in complex systems of various natures [8].

The microscopic mechanism of the PLC effect is understood rather well. It is usually attributed to dynamic strain aging of dislocations, i.e., their repetitive pinning and unpinning from solute clouds [9,10]. The collective unpinning of dislocations leads to jumps in the plastic strain rate and, therefore, abrupt stress relaxations under conditions of loading with a constant strain rate. Such behavior occurs in a bounded domain of strain rate and temperature corresponding to comparable ranges of the solute diffusion time and the waiting time for thermal activation of the dislocation motion.

The strain heterogeneity and stress serration patterns, investigated in detail in tensile tests, display some universal features which allowed for a well-known phenomenological classification [11]. For a given temperature, deformation at a low applied strain-rate $\dot{\varepsilon}_a$ is characterized by a repetitive brief occurrence of localized deformation bands that are almost uncorrelated with each other and result in deep stress serrations below the overall stress level (type *C* behavior). An increase in $\dot{\varepsilon}_a$ reinforces correlation of the deformation bands. Intermediate strain rates are associated with sequences of bands occurring in a correlated manner along the tensile axis, referred to as hopping or relay-race propagation (type *B*). Such bands give rise to series of oscillations around the average stress level. Finally, quasi-continuous propagation of deformation bands along the specimen is observed at high strain rates (type *A*). Repetitive stress rises required to nucleate the bands and followed by abrupt falls to the general stress level are a signature of this type. The periods of band propagation are often caricaturized as corresponding to smooth plastic flow. More exactly, type *A* deformation curves are characterized by stress fluctuations of "all sizes", described by scale-free power-law statistical distributions [8]. Such statistics reflect an avalanche-like character of the relevant processes and are usually interpreted in terms of self-organized criticality (SOC) [12]. Furthermore, the distributions of stress drops are sensitive to the transitions between the above-described types of behavior [8,13–16]. Namely, when $\dot{\varepsilon}_a$ is decreased, the statistics change from scale-free dependences to histograms with a characteristic scale. Therewith, rather complex histograms were observed for type *B* behavior and attributed to deterministic chaos [8]; close-to-Gaussian distributions were found for type *C* serrations [14]. The non-uniqueness of the statistical properties of serrations series is also confirmed by other statistical methods [15,17,18].

Recent investigation of the PLC effect in binary AlMg alloys using acoustic emission (AE) technique brought evidence that the deep stress serrations do not correspond to single dislocation avalanches but are caused by chaining of many avalanches similar to those observed during smooth intervals between serrations [15,16]. As in pure materials characterized by smooth deformation curves [3,19], the AE amplitudes are distributed according to a power law in all deformation conditions, thus corroborating the hypothesis of SOC as a mechanism of collective processes in plasticity. The occurrence of characteristic scales of stress serrations at low strain rate was interpreted in [15,16] as stemming from a competition between SOC and the phenomenon of synchronization also well-known for complex dynamical systems [20][1].

Although the investigations on such model alloys allowed for understanding the origin of the complexity of the PLC effect, the above description does not embrace the entire range of observed behaviors. The binary alloys are not used for practical purposes, mostly because of their low strength. The strengthening is usually based on the formation of strong obstacles to the dislocation motion by adding precipitates and/or refining the polycrystalline grain structure through severe plastic deformation. It occurs that behavior of such alloys may strongly deviate from the above picture. There exist few literature data on the influence of an ultra-fine grain (UFG) structure on the PLC effect. Moreover, the reported results vary from inhibition of serrations, e.g., in dilute AlMg alloys or similar alloys with coarse dispersion particles [23–25], to their intensification, especially, in alloys with nonshearable nanosize precipitates [26–28]. As far as the effect of precipitates is concerned, the literature is abundant, but it mostly deals with the conditions of plastic instability, the type of serrations, and so on, without examining the changes in the dynamical mechanisms of plastic instability (see review [29] and references therein). Recent investigations of the PLC effect in AlMg alloys with Zr and Sc additives in coarse-grained (CG) and UFG conditions testified that these factors can promote one or another type of spatiotemporal patterns of the PLC effect and, therefore, have an impact on the mechanisms of the collective behavior of dislocations [28,30]. The present paper aims at investigating the effect of fine microstructure elements, such as nanosize precipitates and/or submicron grains, on the self-organization of dislocations. With this purpose in view, the statistical analysis of stress serrations and the accompanying AE is applied to an AlMgZrSc alloy with different grain size.

---

[1] Similar duality between the statistics relevant to the macroscopic scale of stress serrations [21] and mesoscopic scale of electronic response to plastic deformation [22] was observed for low-temperature jerky flow caused by thermomechanical instability. This similarity testifies to general mechanisms governing collective dynamics of dislocations for various microscopic mechanisms of plastic instability.

## 2. Experiment and data analysis

The details of the material preparation and the resulting microstructure were described in Ref. [31] and will be briefly outlined here. The CG alloy with the chemical composition Al–6%Mg–0.35%Mn–0.2%Sc–0.08%Zr–0.07%Cr (wt.%) was manufactured by semi-continuous casting and homogenized for 12 h at 360°C. It resulted in an isotropic polycrystalline structure with ~22 μm equiaxed grains and dislocation density of $3\times10^{12}$ m$^{-2}$. The major secondary phase consisted of uniformly distributed coherent $Al_3$(Sc,Zr) dispersoids about 10-15 nm in diameter. Some sparse plate-like or round-shaped incoherent $Al_6$Mn dispersoids with an average thickness of 25 nm were found within the grains and some incoherent $Al_3$(Sc,Zr) particles 40 nm in size appeared on the grain boundaries.

To produce the UFG material, 20 mm × 20 mm × 110 mm rods were processed by Equal Channel Angular Pressing (ECAP) at 320°C to a total strain of ~ 12, according to route $B_C$ [32], in a die with a square cross-section and the channel inner angle of 90°. This procedure resulted in a virtually uniform polycrystalline structure consisting of grains about 700 nm in size, with the recrystallized volume fraction exceeding 90 %, a largely random distribution of misorientations, and the fraction of high-angle boundaries above 80 %. Lattice dislocations pinned by precipitates were revealed in most of the grains. Their density was increased by an order of magnitude but remained relatively low, about $\sim 4\times10^{13}$ m$^{-2}$. The distribution of the second phase particles was qualitatively unchanged (cf. [33]).

The key points of the mechanical testing, the AE measurements, and the statistical analysis of stress serrations and acoustic events were similar to those applied to investigation of binary AlMg alloys [15,16]. Tensile specimens with a dog-bone shape and a gage section of 35 × 7 × 3 mm$^3$ were cut from the ingots, mechanically polished with SiC papers up to 2400 grit, and tested at room temperature and constant grip velocity corresponding to $\dot{\varepsilon}_a$ in the range from $3\times10^{-5}$ to $4.3\times10^{-2}$ s$^{-1}$ with regard to the initial specimen length.

A piezoelectric transducer with a frequency band of 200–900 kHz was clamped to the greased surface of the specimen just above the gage length. The AE captured by the transducer was preamplified by 40 dB in a 10–900 kHz frequency band and recorded using a Euro Physical Acoustic system. The main intricacy in the AE techniques concerns the convention on the criteria applied to individualize meaningful events (hits). Whereas the beginning of the event is defined in an obvious way as the instant when the amplitude of the sound oscillations exceeds the preset noise level ($U_0 = 25$ dB in the present study), its termination is less evident because it depends on the choice made to handle "aftershocks" and unwanted echo signals [34]. In the present work, a standard procedure based on two preset times, namely, a hit definition time (HDT) and a hit lockout time (HLT), was used [35]. First, a hit can contain silent intervals not longer than HDT. In other words, it is considered to come to an end if starting from this instant, the sound oscillations remain below $U_0$ during at least HDT. In addition, the recording apparatus remains in an idle state during HLT after having recorded a hit. The price paid for these precautions is a possible loss of meaningful events and, therefore, depletion of the available statistics. However, it was shown that the AE amplitude distributions are quite robust with regard to both the size of the statistical sample [36,37] and the variation of $U_0$, HDT, and HLT [37]. In the present work, two values of HDT were utilized, 100 and 300 μs. The results of the statistical analysis were similar in both cases. As these values are rather high, so that the aftershocks and the sound reflections are likely to be included into the recorded hit, a relatively small HLT equal to 40 μs was selected. One more characteristic parameter, the peak definition time (PDT), which is applied to detect the peak amplitude of the event, $U$, was equal to half HDT [35]. To avoid confusion hereinafter, the notation $A$ is used for the peak amplitude after conversion from the logarithmic to linear units.

Several rules of thumb tested and adopted in the previous studies [8,15] were applied to the statistical analysis:
- In the case of AE, the analyzed variable was the squared amplitude $I = A^2$ that reflects the energy dissipated in the deformation process, as argued in [38].
- The material work hardening may result in a slow evolution of AE and stress serrations during the test. For this reason, the distributions were calculated over intervals of approximately constant range of variation of the studied variable. In addition, the stationary character of the statistics was verified by varying the length of the intervals.

- The above precaution is less of a constraint in the case of stress serrations, the evolution of which often consists in a simple growth of their average amplitude, without effect on the characteristic patterns. This increase can usually be taken into account through normalization of the stress-time curve $\sigma(t)$ by a smooth function obtained through running-average or polynomial fitting [8]. The magnitudes of stress drops on the normalized deformation curve, which are designated $\Delta s$ hereinafter, are then analyzed.

- Histograms were calculated using data rescaled by the average value of the studied quantity, $X/<X>$, where $X$ means either $I$ or $\Delta s$. This approach avoids arbitrariness in the choice of the bin size by utilizing a unique bin for all calculations. It also facilitates the comparison of the statistical distributions by superposing probability density functions (PDF) obtained for different variables or for the same variable in different conditions.

- Although PDF dependences allow for an intuitively clear interpretation of different behaviors [39], a correct evaluation of power laws is a delicate task, and more exact methods were suggested recently [40,41]. Nevertheless, the comparison of estimates of the power-law exponents obtained by different methods showed that direct calculation of PDF renders reliable results for the PLC effect, providing that the low probability of rare large events is handled accurately [37,42]. The method of bin coupling was used for this purpose. The simple algorithm consisted in grouping initially equal-sized bins by adding the right-hand neighbors until gathering at least five events in each of the resulting bins.

## 3. Results

### 3.1. Stress serrations

Figure 1 compares examples of engineering deformation curves, $\sigma(\varepsilon)$, for samples in CG and UFG microstructure conditions. The overall strain hardening behavior illustrates one of the interests of such alloys: The CG material has a similar ductility but a higher strength than typical binary alloys that usually display ultimate tensile stress below 300 MPa (see, e.g., [14,15]); moreover, ECAP additionally increases the strength of the material without considerably degrading its ductility.

The curves obtained in the cast conditions present signatures of the main types of behavior described above for binary alloys. More specifically, the serration patterns are somewhat atypical at middle strain rates where a mixture of signatures of all three types can be recognized. Anyhow, the overall changes in the patterns produced by $\dot{\varepsilon}_a$ variations correspond to the habitual sequence of transitions between the common types. Curve $1$ demonstrates rather proper type $A$ serrations over a large portion of the curve, which become superimposed with relatively small (about 2 MPa) type $B$ oscillations at large strains. Such $A/B$ transition upon strain hardening is well known in the literature (e.g., [8]). Well-developed type $B$ behavior characterized by deep serrations (about 10 MPa) becomes preponderant at intermediate strain rates (curve $2$) and finally, yet deeper type $C$ serrations occur close to $10^{-4}$ s$^{-1}$ (curve $3$). In all cases, the onset of instability requires a certain critical strain [14]. The smooth character of curve $4$ shows that the low boundary of the strain-rate domain of plastic instability was attained.

Figure 1(b) illustrates that ECAP strongly modifies the appearance of the instability. Jerky flow is observed in the entire strain rate range studied in this work. It occurs immediately upon the yield stress and is dominated by deep type $B$ or type $C$ serrations, while crude type $A$ patterns superimposed with type $B$ oscillations are only discernible at the highest strain rate of $4.3\times10^{-2}$ s$^{-1}$. Rather proper type $B$ patterns occur at $10^{-3}$ and $1.4\times10^{-2}$ s$^{-1}$, as illustrated by curve $2$ for the latter $\dot{\varepsilon}_a$ value. Type $C$ behavior corresponds to the strain rate of $1.4\times10^{-4}$ and $3\times10^{-5}$ s$^{-1}$. The closer view at type $C$ patterns reveals some peculiarities essential for further discussion. Usually, such serrations remain pronounced and even become reinforced during deformation. In contrast, the data of Fig. 1 testify to smoothening of the deformation curves so that the serrations are gradually depressed and seem to give place to behavior resembling type $A$. Such a transition goes in the opposite direction regarding the known strain-hardening effect that corresponds well to the trends described in Fig. 1(a) for the cast alloy.

The statistical distributions of amplitudes of stress serrations in the CG material qualitatively agree with those reported for binary alloys [13–15,43]. Curve $1$ in Fig. 2 demonstrates a power-law PDF, $P(\Delta s) \propto \Delta s^{-\alpha}$, at $\dot{\varepsilon}_a = 1.4 \times 10^{-2}$ s$^{-1}$. This behavior agrees with the SOC hypothesis generally discussed in the case of type *A* serrations. Moreover, such scale invariance covers four orders of serrations magnitude, although it is usually observed over shorter intervals, about two orders of magnitude or less. The values of critical exponent $\alpha$ found for different samples deformed at $\dot{\varepsilon}_a \geq 10^{-3}$ s$^{-1}$ lie within the interval between 1 and 1.8 commonly reported in the literature. Deep type *B* or type *C* serrations give rise to a characteristic scale. At the same time, as was recently pointed out [14,15], the amplitude distributions obtained at slow deformation are often bimodal and contain a small-scale branch obeying a power law. This branch corresponds to irregular fluctuations with small amplitudes below 1 (or several) MPa which are usually disregarded in the literature on the PLC effect. Curve *2* in Fig. 2 represents such a $P(\Delta s)$-dependence in log-log coordinates for the lowest strain rate at which plastic instability was observed in CG material. The rare deep type *C* serrations correspond to two dots at the right edge of the PDF. The amplitudes of small serrations display a power law over more than one order of $\Delta s$, with saturation at low values which is likely due to the stress resolution limit. However, in contrast to the data for a binary AlMg alloy with similar Mg content, for which $\alpha$ in the range of 1–1.5 was found at low strain rates [15], the investigated alloy is characterized by $\alpha$ about 2.5 (see figure caption). Since the higher $\alpha$ means higher probabilities of smaller events relative to larger ones, this difference can be interpreted as a tendency to less correlated deformation processes in the alloy containing precipitates.

The statistical analysis of stress serrations in the UFG material corroborates the above-discussed observations of changes in their morphology. Indeed, convincing scale-free behavior was only detected at low strain rates corresponding to apparent type *C* behavior, $\dot{\varepsilon}_a = 1.4 \times 10^{-4}$ s$^{-1}$, as illustrated by curve *3* in Fig. 2, whereas histograms with a maximum reflecting a characteristic scale were obtained at medium and even high rates. It is noteworthy that scale-invariant behavior represented by curve *3* covers almost the entire range of the observed amplitudes, so that the probability of deep drops only slightly deviates upward from the extrapolated straight line. The scatter between various samples deformed with the same strain rate prevents making a reliable comparison between the power-law dependences found at low $\dot{\varepsilon}_a$ in the UFG and CG materials. Nevertheless, some tendency to lesser $\alpha$-values in the UFG state for the same $\dot{\varepsilon}_a$ can be noticed, as illustrated by curves *2* and *3*.

Figure 3(a) presents an example of a peaked histogram, typical of type *B* serrations (cf. [8]), for one of the high strain-rate values, $\dot{\varepsilon}_a = 5 \times 10^{-3}$ s$^{-1}$. More precisely, the histograms obtained in the fastest tests, at $4.3 \times 10^{-2}$ s$^{-1}$, reveal the common tendency to power-law distributions at fast deformation. Indeed, the example of Fig. 3(b) manifests a roughly monotonically descending dependence. However, even then, the seeming transition to scale-free behavior is far from being completed.

*3.2. Acoustic emission*

Figure 4 gives an example of evolution of AE during a mechanical test. The count rate traced in Fig. 4(b) represents the average AE activity that quantifies the frequency with which the acoustic signal crosses the threshold $U_0$. The overall evolution shown in the figure is typical of most materials [44] and was observed in all experimental conditions in the present work. The AE occurs during elastic deformation, reaches maximum at the transition to macroscopic plasticity (yield limit), and further decreases in the course of deformation. The presence of AE during the elastic stage is usually explained by micro-plasticity [45]; the activity growth during the elastoplastic transition is explained by an intense multiplication of dislocations; and the further decrease is ascribed to the accumulation of obstacles to the motion of dislocations. The last factor can also justify the depression of the AE intensity at large strains, which can be recognized in Fig. 4(c) resolving series of amplitudes of individual hits. When the strain rate is varied, such series display diverse patterns described in detail elsewhere [15,16].

Several observations are noteworthy here. The amplitudes of the hits occurring at the instants of stress drops usually do not stand out from those generated during smooth plastic flow. For example, the prominent event shown by an arrow in Fig. 4(c) does not correspond to a stress serration. On the other hand, the serrations are correlated with bursts in the event durations $\tau$, as marked by rectangles in Figs.

4(a) and 4(d). Such bursts were explained in [15,16] in terms of synchronization of dislocation avalanches in the sense of the chaining of many acoustic events, each one reflecting the occurrence of a dislocation avalanche, so that they are recorded as a single event with a long duration. In comparison with this behavior, which is characteristic of type *C* serrations occurring downward from the general stress level, bursts in both *U* and $\tau$ are observed during stress rises accompanying type *B* and type *A* curves. Such rises mark the stages of nucleation of a new deformation band after the material has been work-hardened by the passage of the previous band (type *A*) or the previous sequence of bands (type *B*) through the entire gage length. These observations apply to both CG and UFG alloys. The major differences between these two cases are that for the same strain rate, the AE activity is always essentially weaker (two to ten times) and the maximum amplitude can be much higher in the UFG alloys. Typically, *U* reaches 60–70 dB in CG state and 60–90 (sometimes up to 100 dB) in the UFG state.

As with the results for binary AlMg alloys [15,16], the statistical analysis revealed power-law distributions of AE intensity, $P(I) \propto I^{-\alpha_{AE}}$, in all experimental conditions. Unexpectedly, the investigated alloys demonstrated a qualitatively different effect of strain on the critical exponent $\alpha_{AE}$, as compared with that reported for binary alloys. Figure 5 presents examples of power-law dependences determined over two strain intervals for a CG alloy deformed at $\dot{\varepsilon}_a = 10^{-3}$ s$^{-1}$. It cannot escape ones's attention that the respective slope is considerably lower at large strain, while an opposite trend was found in [15]. Such strain effect was observed in most tests; Fig. 6 shows deformation curves with the evolution of $\alpha_{AE}$ for one specimen of each kind and three strain rates. The strain dependences of $\alpha_{AE}$ found for these and other samples are quite diverse, most likely because of the influence of the initial microstructure that may strongly vary between samples subjected to complex treatments. Nevertheless, several common features can be identified. The exponent usually takes on relatively low values during the initial quasi-elastic deformation. More specifically, it is close to the value of 2 which approaches typical exponents reported for various materials in the absence of dynamic strain aging and is similar to the exponent found in cyclic tension-compression of pure aluminum [19]. As in [15], $\alpha_{AE}$ strongly grows in the region of the elasto-plastic transition. However, while it was found to stabilize or even continue growing during deformation of binary alloys, Fig. 6 demonstrates a clear tendency to a decrease in $\alpha_{AE}$ with further deformation. An exception to this rule is shown by solid lines in Fig. 6(a). It corresponds to a CG specimen deformed at $\dot{\varepsilon}_a = 1.4 \times 10^{-2}$ s$^{-1}$, which demonstrates a pattern of macroscopic serrations very similar to that observed in binary alloys.

Two more observations should be mentioned. First, for large enough strain, $\alpha_{AE}$ is usually lower in the UFG material than in the CG alloy, which agrees with the similar relationship for the statistics of stress serrations (see Sec. 3.1.) Finally, it is noteworthy that the maximum $\alpha_{AE}$ observed for both investigated alloys is as high as 4, while it did not exceed 3 for binary alloys.

## 4. Discussion and conclusions

The stress serration patterns and statistics bear evidence to specific features of the investigated materials regarding the transitions between PLC behavior types. The CG alloy displays an unusual proneness to power-law distributions over the entire $\dot{\varepsilon}_a$-range (Fig. 2). This result agrees with the recent observation of the predominance of the PLC band propagation (type *A* behavior) for the same material [30]. The UFG alloy displays even less common features, so that scale-free statistics are only observed at the lowest strain rate, in consistence with the unusual transition from type *C* to type *A* serrations after some deformation. Only a weak trend to power-law statistics can be recognized at high $\dot{\varepsilon}_a$, in agreement with the incomplete *B* to *A* transition in the serration patterns. These distinct differences from behavior of binary alloys testify that the microstructure can strongly affect the self-organization of dislocations. The analysis of the AE statistics can shed additional light on the role of the microstructure. To interpret the observed peculiarities, let us first summarize some of the overall suggestions stemming from the AE investigations.

The observation of a persistent power-law character of the AE statistics for the studied alloys confirms the general conjecture of an inherently scale-free avalanche-like dynamics of dislocations, which applies to both unstable plastic deformation and macroscopically stable flow and is usually attributed to the SOC mechanism [3-8,15,16,19]. The general validity of this conjecture is confined from above by a mesoscopic scale pertaining to the AE and, perhaps, low-amplitude stress serrations in the case of the PLC effect. As discussed in [46], the smoothness of the macroscopic deformation curves of most materials makes one suggest the existence of inherent factors limiting the size of the dislocation avalanches, in particular, intrinsic lengths related to the microstructure and the crystallography of the dislocation glide. Although it is tentative to attribute the PLC serrations to formation of extremely powerful dislocation avalanches in the presence of the dynamic strain aging, the above-said limitations must also apply to the macroscopically unstable deformation. This assumption follows from the observation of a similar AE intensity range at the instants of deep serrations and during intervals of smooth plastic flow at low enough strain rate (Fig. 4). The concomitant observation of the event duration bursts has led to a conjecture that the macroscopic stress serrations occur through dense clustering of dislocation avalanches of the same nature as in the absence of the PLC instability [15,16]. Furthermore, the transitions between various types of macroscopic behavior with the strain-rate variation can be understood as a competition between SOC and synchronization phenomena, as predicted in some generic models of complex dynamical systems [20]. It can be outlined as follows. The strain and stress heterogeneity generated by a macroscopic instability is smoothed out during the subsequent reloading. This homogenization takes place through plastic deformation processes producing negligible effects on the deformation curve, namely, individual dislocation avalanches responsible for discrete AE events (Fig. 4) and the motion of individual dislocations and small dislocation pile-ups, giving rise to a continuous acoustic noise [44]. At slow deformation, a high level of uniformity is attained during the large reloading time between serrations. Therefore, a dense sequence of dislocation avalanches can be triggered when the instability condition is reached again. In this situation, the size and duration of the stress serrations are mainly determined by the elastic unloading of the deformation machine, and their distributions display a bell shape corresponding to fluctuations from the ideal case of periodic relaxation oscillations [47] predicted in early models of the PLC effect [48]. In the opposite limit of fast deformation, the local strain field is constantly highly heterogeneous and gives rise to deformation processes on all scales (below that of the relaxation oscillations), in agreement with the power-law statistics of type *A* serrations in a bounded scale range.

At a given strain rate, the effectiveness of relaxation processes depends on the material microstructure. It can be expected that the statistics of the discrete AE may provide quantitative information on the microstructure effects. Table 1 summarizes the data obtained in the present work and the literature data on the power-law exponents determined from the distributions of stress serrations and AE for various materials. The first analyses of the AE accompanying plasticity were mostly performed on single crystals with an anisotropic *hcp* structure (ice, Cd, Zn, etc.) [3]. Plastic deformation of such materials is mostly constrained to one slip system. The results obtained led to an assumption of universal power-law behavior with $\alpha_{AE} \sim 1.5$ for the parameters characterizing the dissipated energy, notably, $A^2$. Considering that the mechanical work associated with a strain jump performed at a constant stress is proportional to its size, this assumption was also confirmed by a similar exponent value for statistics of strain jumps during deformation of micropillars [6]. However, the data obtained for bulk polycrystalline ice samples [3] and materials with an isotropic cubic structure (Al, Cu, etc.) [19] testified that the power law can be sensitive to the glide crystallography and the microstructure. The variation of $\alpha_{AE}$ between 1.5 and 2 observed in [19] was attributed to a stochastic factor caused by the multiple slip leading to forest hardening, formation of dislocation structures, etc., therefore, reducing the probability of large avalanches.

Another general explanation of such a variation, in terms of the dynamics of tilted slip bands considered as ellipsoidal avalanches of dislocations, was proposed in [49]. However, this model only applies to the linear work-hardening stage that is not observed in the case of the PLC effect. As far as the stochastic factor is concerned, it might acquire more importance in alloyed materials because of the additional solute hardening. Indeed, this conjecture was used to explain high $\alpha_{AE}$-values found for binary

AlMg alloys [15,16]. More exactly, $\alpha_{AE} \sim 2$ was found during quasi-elastic stage, in agreement with the data for pure Al [19]; it increased up to 3 upon the transition to macroscopic plastic flow leading to accumulation of obstacles to the dislocation motion. These arguments are corroborated by the present data displaying even higher upper bounds of $\alpha_{AE}$ during microplastic flow and elastoplastic transition in the materials with complex initial microstructures (Table 1; Figs. 5 and 6). Indeed, both precipitates and grain boundaries are effective obstacles to the dislocation motion.

At the same time, the randomization of the dislocation motion cannot explain the strong decrease in $\alpha_{AE}$ back to the value of 2 at larger strains (see Figs. 5 and 6). In addition, the tendency to lower $\alpha_{AE}$-values in the UFG state than in the CG material also contradicts this model. To interpret the entirety of data, a competition of several factors should be considered. Indeed, while arresting mobile dislocations on obstacles breaks the correlation of deformation processes, the resulting concentration of internal stresses would act in the opposite sense and enhance the dislocation avalanches through nucleation and remobilization of dislocations in the same grain and initiation of dislocation glide in the neighboring grains. The latter assumption was applied to explain relatively low $\alpha_{AE}$ in polycrystalline ice (see Table 1) [3]. Besides, grain boundaries may play a double role, assisting either the local stress concentration because of the dislocation pinning, or its relaxation because of the dislocation drainage and nucleation of new dislocations. For example, it may explain the globally weaker AE activity in UFG alloys. The real behavior of complex alloys may thus result from a subtle balance between all these factors.

This framework also sheds light on the differences between the power laws determined from the AE and stress serrations in the same conditions (Table 1). Let us first recall that at low strain rate, synchronization of dislocation avalanches leads to deep serrations with the size distribution tending to the Gaussian shape, as observed in binary alloys [8,13,14]. The data obtained in the present work testify that additional strong obstacles to the dislocation motion weaken the synchronization and result in power-law distributions in a relatively large range of serration amplitudes (Figs. 2 and 3). Therewith, the value of $\alpha$ agrees quite well with $\alpha_{AE}$ in the strain interval corresponding to the serrated flow, thus confirming that both kinds of "events", the stress serrations and the accompanying AE, reflect the same dislocation processes. Despite the stronger synchronization effect in binary alloys, power-law distributions can be detected for the subset of small stress serrations at low strain rates. However, the respective $\alpha$-value is considerably lower than $\alpha_{AE}$ (Table 1). This difference is most likely due to a low frequency band of the load cell ($\leq 500$ Hz), so that dense sequences of avalanches resolved by the acoustic system are detected as single events with the amplitude determined by the summary effect of many avalanches. The same argument can explain inequality $\alpha < \alpha_{AE}$ in the high strain-rate tests, even if these conditions correspond to a distinct dynamical regime and are characterized by scale-free distributions of both AE and stress serrations. In this case, the high frequency of dislocation avalanches, reflected in a very strong AE activity [16], is simply due to the need to assure the high plastic strain rate.

In summary, the results of AE investigations suggest a unique approach to interpretation of various serration patterns and, more generally, spatiotemporal behavior of the PLC effect (cf. [30] regarding the deformation band kinematics). Although the fast and slow deformation regimes are known to correspond to qualitatively different dynamical modes of unstable flow, the formation of the macroscopic instability and the transitions between different regimes can be rationalized within a general framework considering dislocation avalanches as basic elements. More precisely, the validity of this approach does not only follow from the study of the PLC instability but is also corroborated by the avalanche-like character of smooth plastic flow on the AE scale. In the present work, it has been demonstrated using the example of an AlMgZrSc alloy with two grain sizes that the analysis of the factors controlling the strength of correlation between such avalanches may shed light on some peculiarities of the behavior of applied alloys with complex composition and microstructure.

**Acknowledgments**
The authors are grateful to P. Moll for help in experiments. This work was partly supported by the Center of Excellence "LabEx DAMAS" (Grant No. ANR-11-LABX-0008-01 of the French National Research

Agency) and by the Togliatti State University (Grant No. 14.Z50.31.0039 of the Ministry of Education and Science of the Russian Federation).Agency) and by the Togliatti State University (Grant No. 14.Z50.31.0039 of the Ministry of Education and Science of the Russian Federation).

Table 1

| Characteristic | Conditions | Exponent |
|---|---|---|
| AE | Bulk anisotropic materials [3,19] | ~ 1.5 (single crystals); ~ 1.2 (ice polycrystals)* |
| | Bulk isotropic materials [19] | 1.5÷2 |
| | Binary AlMg alloys [15-16] | 1.8÷2.2 during microplastic flow. Increases up to 2.5÷3.1 at larger strains. |
| | Present study (Figs. 5 and 6) | 1.8÷2.5 during microplastic flow. 2.5÷4 at elastoplastic transition. Decreases down to 1.6÷2.2 at larger strains**. |
| Strain or stress jumps | Micropillars [6] | ~ 1.5 |
| | Binary AlMg alloys [8,13,14] | High $\dot{\varepsilon}_a$: 1.0÷1.5. Intermediate and low $\dot{\varepsilon}_a$: bell-shaped histograms. Lowest $\dot{\varepsilon}_a$: 1.0÷1.5 for the subset of low-amplitude stress fluctuations. |
| | Present study (Figs. 2 and 3) | High $\dot{\varepsilon}_a$: 1.0÷1.8 for CG alloy; incomplete transition to power law for UFG alloy. Intermediate and low $\dot{\varepsilon}_a$: 1.9÷2.5 (in the case of UFG alloy, the power law is only detected at the lowest $\dot{\varepsilon}_a$ and displays a tendency to lower $\alpha$-values than for the CG material). |

* The estimate of $\alpha_{AE}$ for $A^2$ was obtained from the exponent $\alpha_A \approx 1.35$ determined for amplitude distributions [3]: $\alpha_{AE} = (\alpha_A +1)/2$.

** A monotonous increase, as in the case of binary alloys, was observed at the highest strain rate for a CG sample demonstrating pure type *A* serrations (see description in the text).

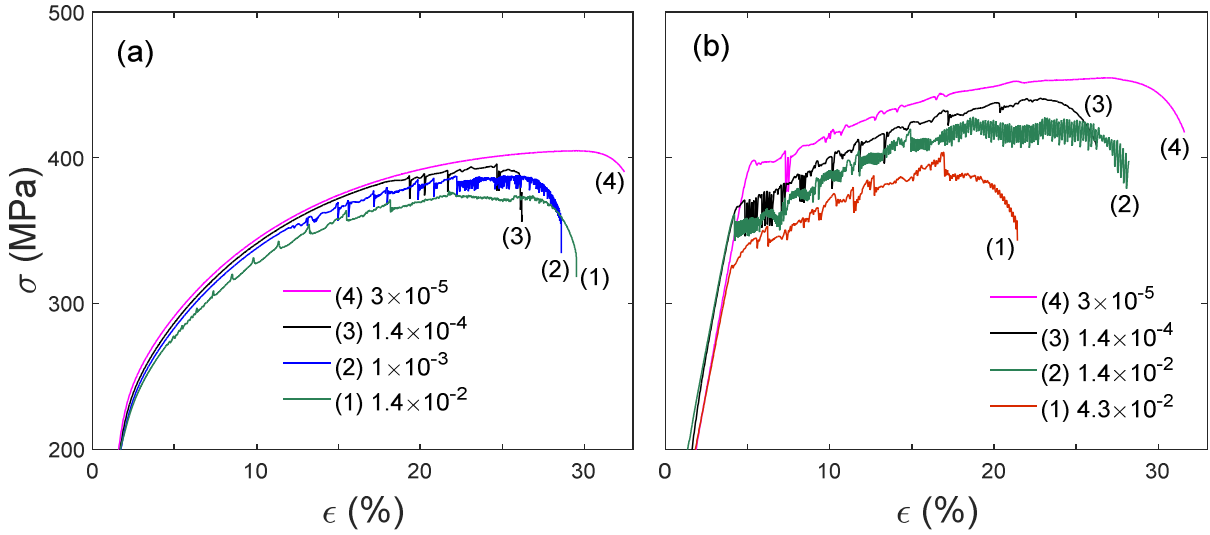

**Figure 1**. Examples of engineering stress-strain curves $\sigma(\varepsilon)$: (a) CG alloy, (b) UFG alloy. The values of the imposed strain rate are given in s$^{-1}$. For clarity, curve *1* in the chart (b) is shifted by 20 MPa downwards.

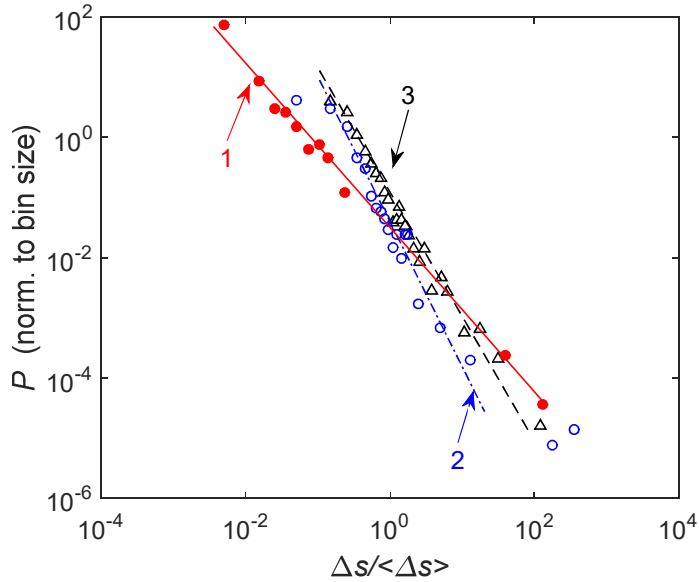

**Figure 2**. Examples of the probability density function $P(\Delta s / <\Delta s>)$ for amplitudes of serrations: 1 - CG alloy, $\dot{\varepsilon}_a = 1.4\times10^{-2}$ s$^{-1}$; the least-squares estimate of the slope and the corresponding root-mean-square error render $\alpha \approx 1.4\pm0.1$. 2 - CG alloy, $\dot{\varepsilon}_a = 1.4\times10^{-4}$ s$^{-1}$, $\alpha \approx 2.5\pm0.2$. 3 - UFG alloy, $\dot{\varepsilon}_a = 1.4\times10^{-4}$ s$^{-1}$, $\alpha \approx 1.9\pm0.1$.

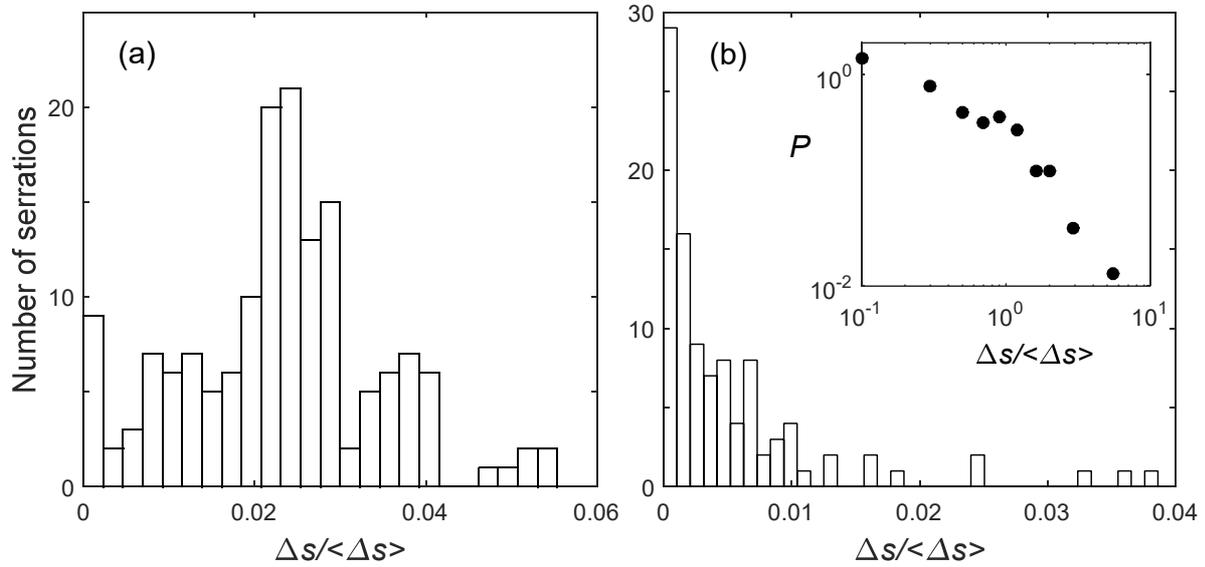

**Figure 3**. Examples of histograms of distribution of stress serration amplitudes for the UFG alloy deformed at high strain rates: (a) $\dot{\varepsilon}_a = 5\times10^{-3}$ s$^{-1}$, (b) $\dot{\varepsilon}_a = 4.3\times10^{-2}$ s$^{-1}$. Insert: the corresponding PDF dependence.

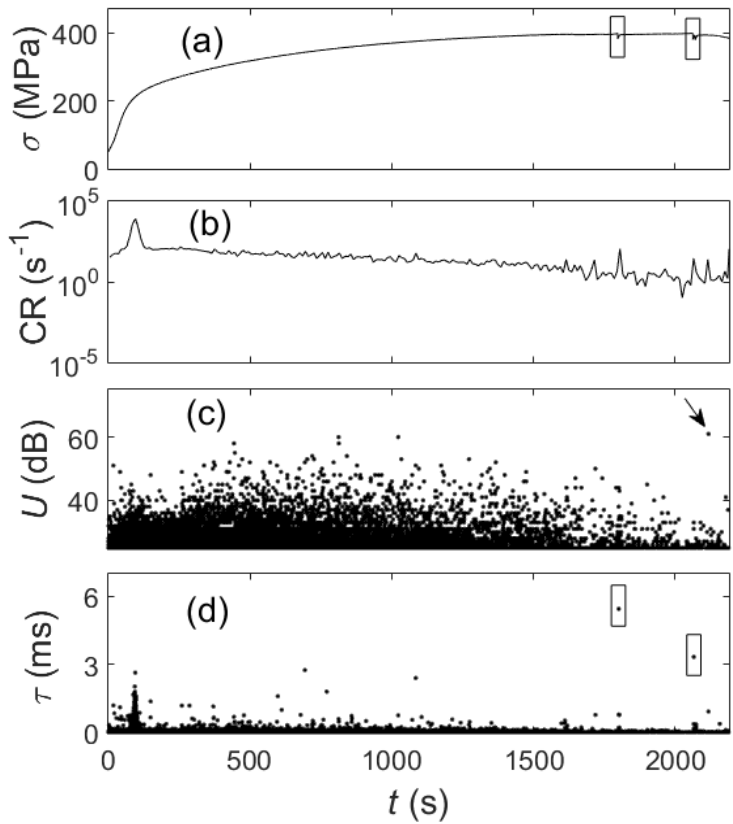

**Figure 4**. (a) Time-stress curve $\sigma(t)$ and the evolution of various characteristics of the accompanying AE: (b) the average count rate (CR) calculated over 10 s time intervals, (c) logarithmic amplitudes $U$ of AE events, and (d) the corresponding durations $\tau$. $\dot{\varepsilon}_a = 1.4\times10^{-4}$ s$^{-1}$. Rectangles in plots (a, d) show two macroscopic stress serrations and the respective bursts in $\tau$. The arrow in plot (c) indicates a high-amplitude event during macroscopically smooth plastic flow.

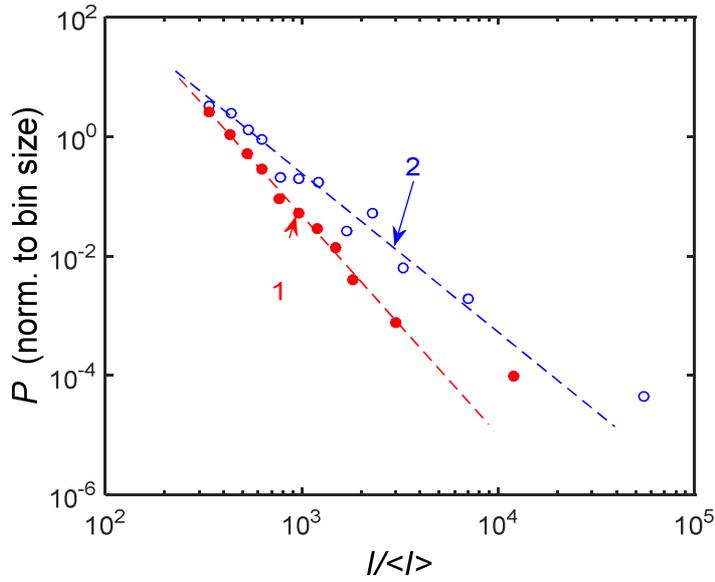

Figure 5. Examples of the probability density function $P(I/\langle I\rangle)$ of the AE intensity for a CG sample deformed at $\dot{\varepsilon}_a = 10^{-3}$ s$^{-1}$. 1 – strain interval from 3% to 6 %, $\alpha_{AE} \approx 3.6\pm0.1$. 2 - strain interval from 15% to 20 %, $\alpha_{AE} \approx 2.4\pm0.1$.

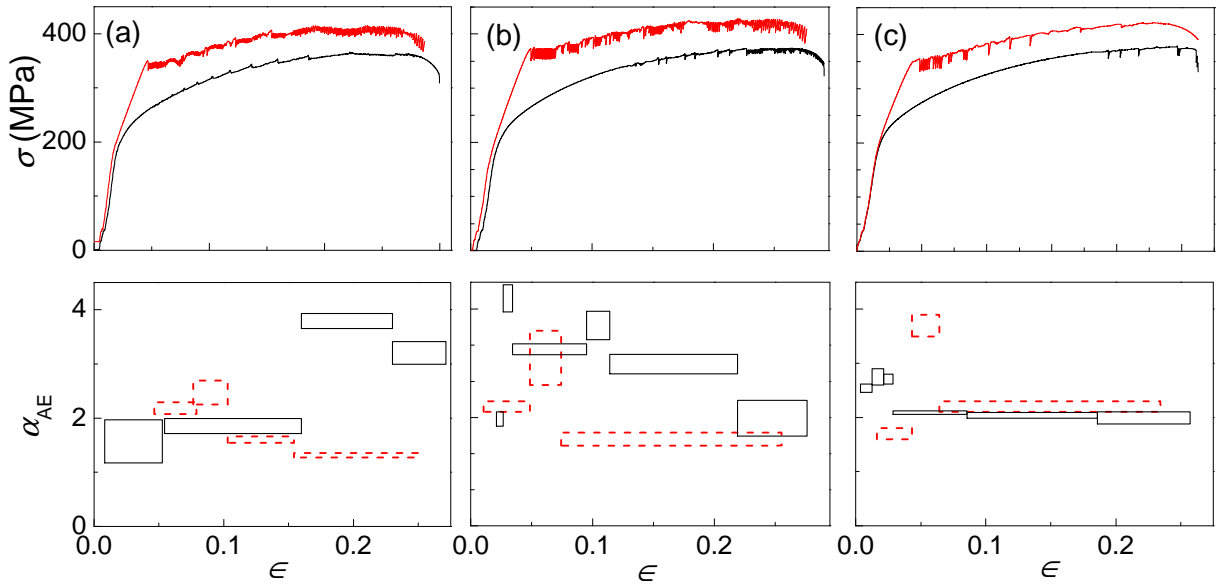

Figure 6. Examples of comparison of deformation curves with evolution of the critical exponent $\alpha_{AE}$ for the AE intensity statistics. (a) $\dot{\varepsilon}_a = 1.4\times10^{-2}$ s$^{-1}$; (b) $\dot{\varepsilon}_a = 10^{-3}$ s$^{-1}$; (c) $\dot{\varepsilon}_a = 1.4\times10^{-4}$ s$^{-1}$. All deformation curves are traced using solid lines: The curves for the UFG alloy pass above their counterparts for the CG alloy (cf. Fig. 1). Rectangles in the bottom plots show the length of the strain intervals over which $\alpha_{AE}$-values were estimated and the corresponding root-mean-square errors. Solid and dashed lines present results of calculation for the CG and UFG alloys, respectively.